\documentclass[aps,psfig,prd,preprintnumbers,amsmath,amssymb,
superscriptaddress,twocolumn,floatfix]{revtex4}

\usepackage{graphics}
\begin{document}
\title{Cosmology with massive neutrinos coupled to dark energy}
\author{A.~W.~Brookfield}
\affiliation{Department of Applied Mathematics and Department of Physics, Astro--Particle Theory $\&$ Cosmology 
Group, Hounsfield Road, Hicks Building, University of Sheffield, Sheffield S3 7RH, 
United Kingdom}
\author{C.~van de Bruck}
\affiliation{Department of Applied Mathematics, Astro--Particle Theory $\&$ Cosmology 
Group, Hounsfield Road, Hicks Building, University of Sheffield, Sheffield S3 7RH, 
United Kingdom}
\author{D.~F.~ Mota} 
\affiliation{Institute of Theoretical Astrophysics, University of Oslo, 0315 Oslo, Norway}
\affiliation{Astrophysics Department, Oxford University, Keble Road, Oxford OX1 3RH, United Kingdom}
\author{D.~Tocchini-Valentini}
\affiliation{Astrophysics Department, Oxford University, Keble Road, Oxford OX1 3RH, United Kingdom}
\date{01 July 2005}
\begin{abstract}
Cosmological consequences of a coupling between massive neutrinos and dark energy are investigated. 
In such models, the neutrino mass is a function of a scalar field, which plays the role of dark energy. The evolution 
of the background and cosmological perturbations are discussed. We find that mass--varying 
neutrinos can leave a significant imprint on the anisotropies in the CMB and even lead to a reduction of 
power on large angular scales. 
\end{abstract}
\maketitle 
The discovery of the accelerated expansion of the universe is a major challenge 
for particle physics (see \cite{riess} for latest results). According to General Relativity, the dynamics of 
the universe is dominated by a new (dark) energy form with negative pressure.
A well-motivated candidate for dark energy is a light scalar field \cite{wetterich,ratra}. 
From the particle physics point of view, however, such light scalar fields are problematic: first of all, why is 
the mass so small and how can this low mass be stabilized against radiative corrections \cite{doran}? 
Secondly: is this field coupled to any other matter form? And if not, why not \cite{rest}? 
Indeed, early papers on quintessence discussed this possibility in detail with models in which 
dark matter is coupled to dark energy \cite{wetterich,wetterich2,amendola}. The discovery of ``dark energy'' clearly 
requires new physics for its explanation. 

In this work we explore the cosmological consequences of an idea recently put forward  in \cite{fardon}. 
According to this idea, dark energy and neutrinos are coupled such that the mass of the neutrinos is a 
function of the scalar field which drives the late time accelerated expansion of the universe. In general, 
the field will evolve with time and, hence, the mass of the neutrinos is not constant (mass-varying neutrinos). 
One of the motivations for such considerations is the question of whether there is a relation between the neutrino mass scale and the 
dark energy scale, that has a similar order of magnitude compared to the detected neutrino mass splittings. In such models the origin 
of the neutrino mass and dark energy are interlinked. Astrophysical and cosmological implications of such models have recently been 
studied in \cite{followups}. Here we study for the first time the transition of coupled neutrinos from the relativistic 
to the non-relativistic regime as well as the dynamics of the dark energy field. 
We also consider how the coupling affects the cosmic microwave background radiation (CMB) and large scale 
structures (LSS). 

The dark energy sector is described by a scalar field with potential energy $V(\phi)$. This potential has to be seen as 
an effective, classical one, since the coupling between the scalar field and the neutrinos can lead to significant 
quantum corrections \cite{doran}, a problem also present in models with dark matter/dark energy interaction \cite{amendola}.
To be specific, in this letter we will choose a standard quintessential potential, namely the exponential potential 
$V(\phi) = V_0 \exp(-\sqrt{3}\lambda\phi/\sqrt{2})$ (in the following we set $8\pi G \equiv 1)$ \cite{wetterich}. 
With this choice, our theory differs from the one proposed in \cite{fardon}. There, the choice of 
potential was such that the mass of the scalar field is much larger than the Hubble parameter $H$ from times before 
big bang nucleosynthesis until today. In contrast, with our choice of potential, the mass of the field will be at 
most of order $H$. For our purposes the neutrinos can be either Dirac or Majorana particles: the details will 
not affect our considerations. The only necessary ingredient is that, according to \cite{fardon}, the neutrino 
mass is a function of the scalar field, i.e. $m_\nu = m_\nu(\phi)$. Here we consider three species of neutrinos 
with the same mass and choose a field--dependence of the form $m_\nu = M_0 \exp(\beta \phi)$, with $\beta={\cal O}(1)$. 
The form of the coupling chosen is well motivated (see e.g. \cite{wetterich,wetterich2}) and has been 
considered in the past in models with dark matter/dark energy interaction \cite{wetterich2,amendola}. 
We point out, however, that results for other potentials and couplings are similar to the ones presented 
here \cite{ourlater}. 

In the cosmological context, neutrinos cannot be described as a fluid. Instead, we must 
solve the distribution function $f(x^i,p^i,\tau)$ in phase space (where $\tau$ is the conformal 
time). We are interested in times when neutrinos are collisionless, and so the 
distribution function $f$ does not depend explicitly on time. Solving the Boltzmann equation, 
we can then calculate the energy density stored in neutrinos ($f_0$ is the background neutrino 
distribution function):
\begin{equation}
\rho_\nu = \frac{1}{a^4} \int q^2 dq d\Omega \epsilon f_0(q), 
\end{equation}
with $\epsilon^2 = q^2 + m_\nu(\phi)^2 a^2$, $a$ is the scale factor and $q^i = a p^i$ is the 
comoving momentum. The pressure is 
\begin{equation}
p_\nu = \frac{1}{3 a^4} \int q^2 dq d\Omega f_0(q) \frac{q^2}{\epsilon}.
\end{equation}
From these equations one can easily derive that 
\begin{equation}\label{neutrinoequation}
\dot\rho_\nu + 3H(\rho_\nu + p_\nu) = \frac{\partial \ln m_\nu}{\partial \phi}\dot\phi(\rho_\nu 
- 3p_\nu)
\end{equation}
(the dot representing the derivative with 
respect to $\tau$). 
This equation is akin to the equation for matter coupled to a scalar field 
in scalar--tensor theories. The equation of motion for the scalar field reads 
\begin{equation}\label{scalarequation}
\ddot\phi + 2H\dot\phi + a^2\frac{\partial V}{\partial\phi} = 
- a^2\frac{\partial \ln m_\nu}{\partial \phi}(\rho_\nu-3p_\nu),
\end{equation}
which can be obtained from the energy conservation equation of the {\it combined} fluid of neutrinos and 
dark energy 
\begin{equation}
\dot \rho_\nu + \dot \rho_\phi  + 3H(\rho_\nu + \rho_\phi + p_\nu+p_\phi) = 0
\end{equation}
and eq. (\ref{neutrinoequation}). From this equation and our choice of $m_\nu(\phi)$, 
the  dynamic of the field is specified by the effective potential 
\begin{equation}
V_{\rm eff} = V(\phi) + (\tilde\rho_\nu -3\tilde p_\nu)e^{\beta\phi},
\end{equation}
where $\tilde\rho_\nu = \rho_\nu e^{-\beta\phi}$ and $\tilde p_\nu = p_\nu e^{-\beta\phi}$ are independent 
of $\phi$. With $V = V_0 e^{-\sqrt{3}\lambda\phi/\sqrt{2}}$, the field value at the minimum of the effective potential 
is given by $\phi_{\rm min} = \lambda^{-1}\ln (\sqrt{3}\lambda V_0/\sqrt{2}\beta(\rho_\nu-3p_\nu))$. Restricting 
to the case $\lambda>0$, the effective minimum only exists for $\beta>0$. 

\begin{figure}
\centerline{\scalebox{0.35}{\includegraphics{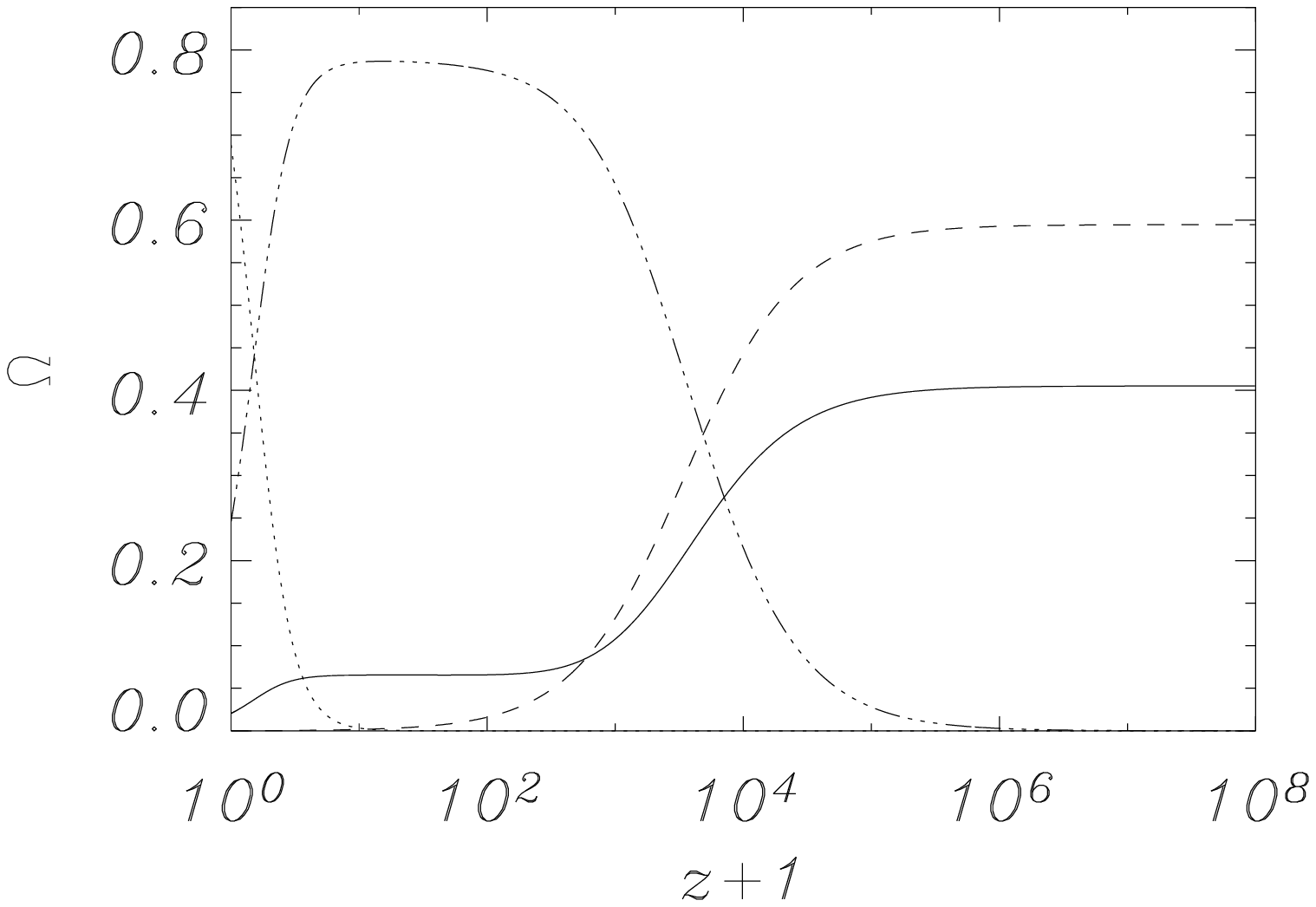}}}
\centerline{\scalebox{0.35}{\includegraphics{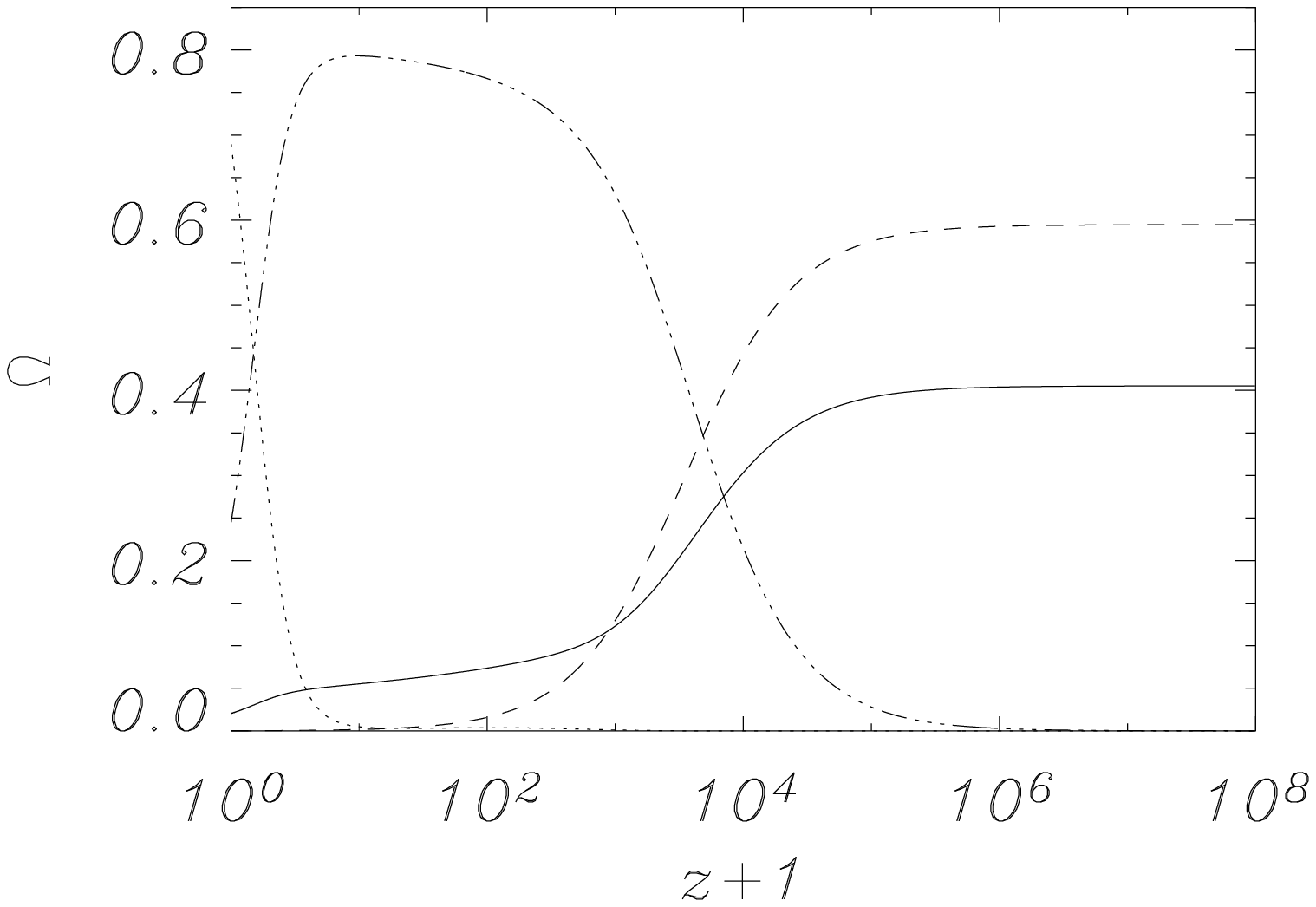}}}
\caption{Background evolution: In the upper panel, we plot the evolution of the density parameters 
for a model with $\beta=0$, $\lambda=1$. In the lower panel the corresponding plot with $\beta=1$ is shown. 
(Neutrinos: solid line, CDM: dot-dashed line, scalar field: dotted line and radiation: dashed line.) 
In all cases, the mass of the neutrinos is $m_\nu = 0.314$ eV today. We are considering a 
flat universe with $\Omega_b h^2 = 0.022$, $\Omega_c h^2 = 0.12$, $\Omega_\nu h^2 = 0.01$ 
and $h=0.7$.}
\end{figure}

With our choice of potential and coupling, the setup is similar to that studied in \cite{domenico}. 
For such a system, a number of critical points have been identified of which only two of these
are compatible with an accelerating universe. However in our work there is the major difference that the dark energy field 
couples to neutrinos rather than cold dark matter (CDM). Furthermore neutrinos have never dominated the dynamics of the universe 
in the past and their equation of state is not constant. *** When the neutrinos are relativistic, the coupling terms containing 
the trace of the neutrino energy momentum tensor are small, but non-zero. In particular, $(\rho_\nu - 3 p_\nu)/\rho_\nu \ll 1$. 
Given that $\dot\phi$ is at most of order $H$ in quintessence models, and $\beta$ is of order one, it is obvious that the second 
term on the left hand side in eq. (\ref{neutrinoequation}) dominates over the coupling term. We have confirmed this by numerically 
solving the Boltzmann equation. The source terms are only significant when the neutrinos become non--relativistic. ***

\begin{figure}
\centerline{\scalebox{0.35}{\includegraphics{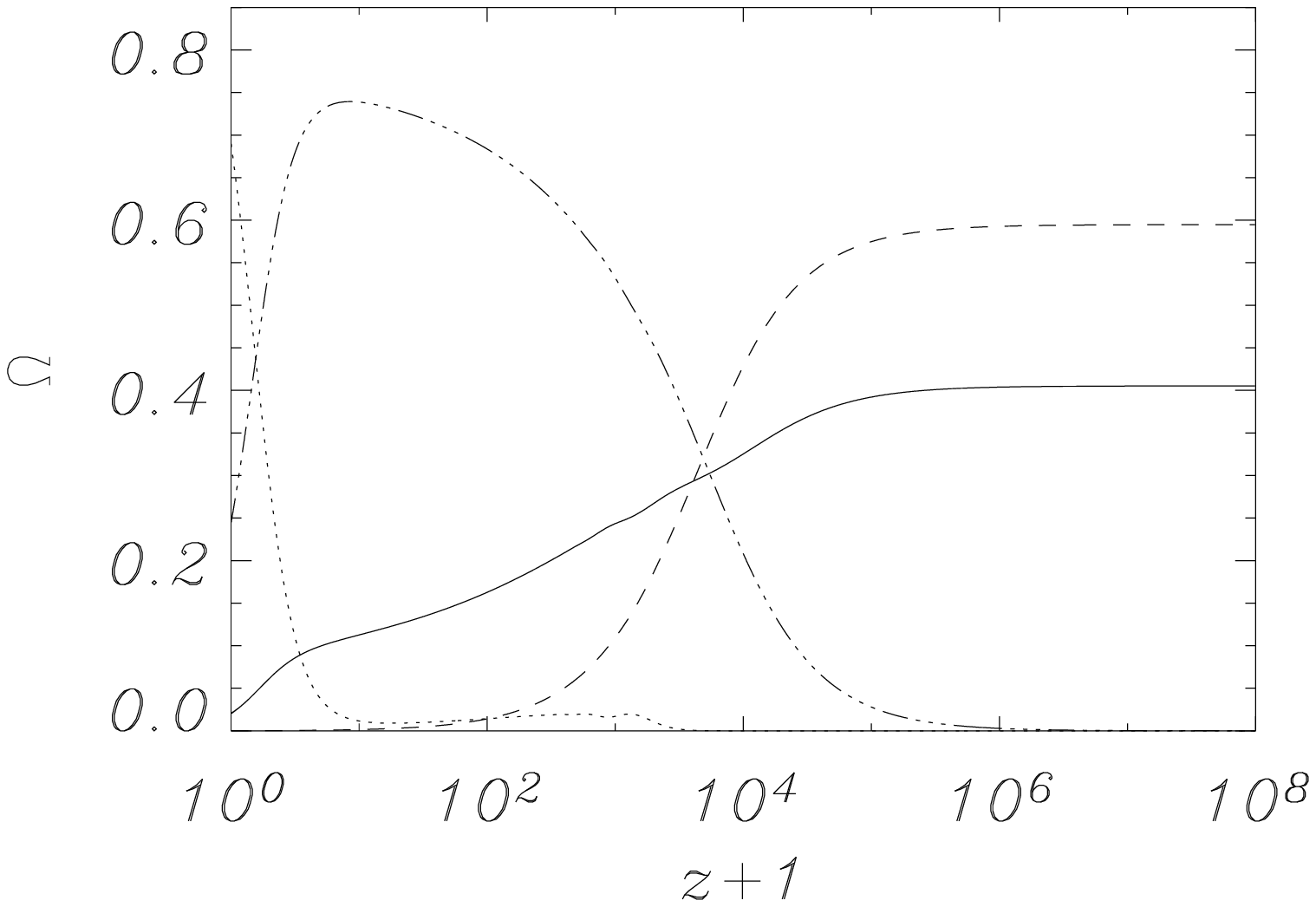}}}
\centerline{\scalebox{0.35}{\includegraphics{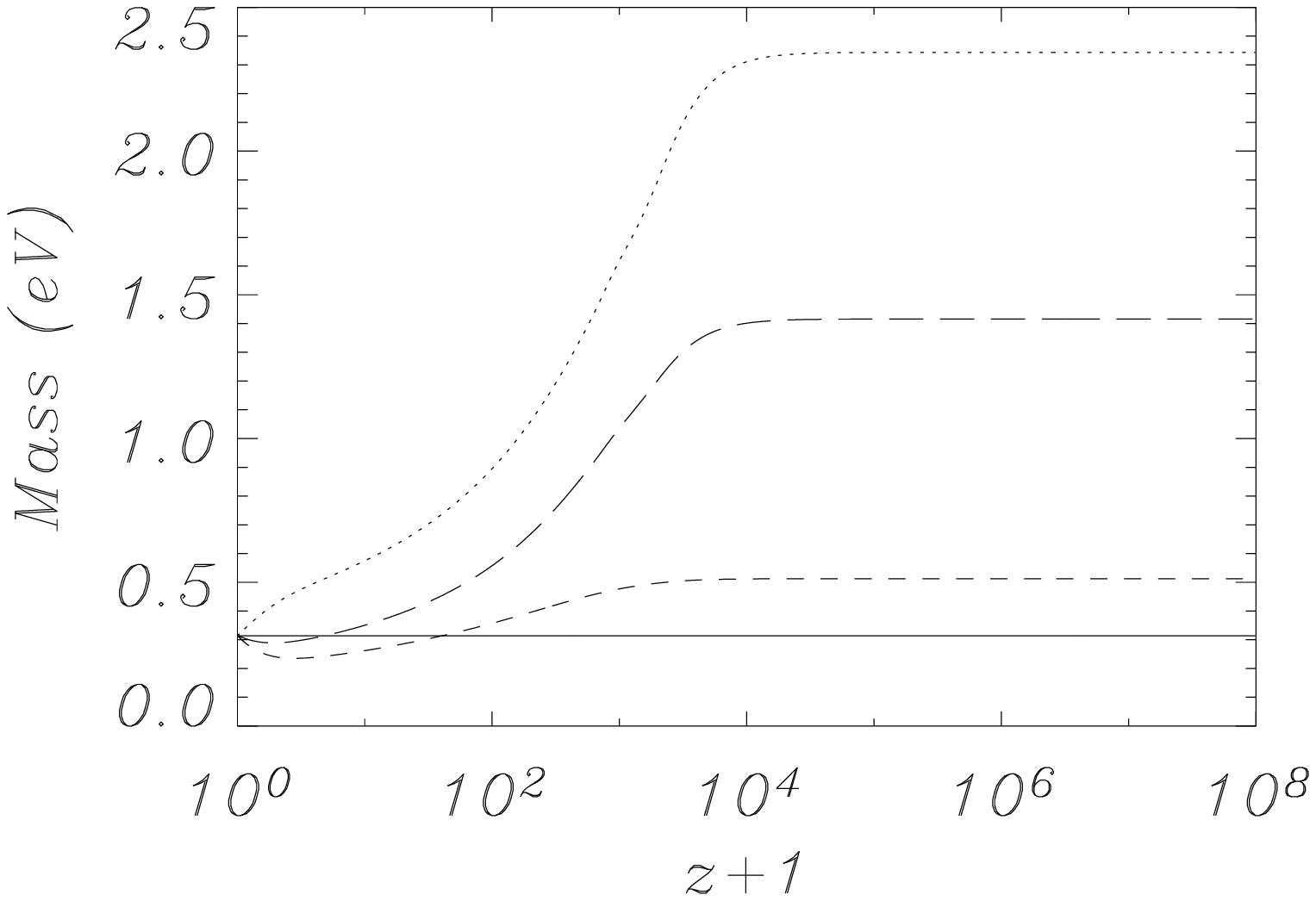}}}
\caption{The upper plot is the same as Figure 1, but 
choosing $\beta=-0.79$ and $\lambda = 1$. The cosmological parameters are chosen as in Figure 1. The lower 
plot shows the evolution of the neutrino mass in the different models (solid line: $\beta=0$, $\lambda=1$; 
short dashed line: $\beta=1$, $\lambda = 1$; dotted line: $\beta=-0.79$, $\lambda = 1$; long dashed line 
$\beta = 1$, $\lambda = 0.5$.)}
\end{figure}

Typically, we find that the system passes through a series of four stages. Firstly, 
when the neutrinos are ultra--relativistic, the field is frozen and the neutrino mass 
is constant. Then, as the neutrinos start to become 
non--relativistic, part of the energy of the neutrinos is transferred to kinetic energy for the scalar field. 
Subsequently, at a temperature close to the neutrino mass, the neutrinos become non--relativistic and begin to scale similarly to 
dark energy but differently compared to dark matter. Here the kinetic energy dominates the dynamics of the scalar field. During 
this time the neutrino mass starts to evolve significantly. Finally, typically at a redshift of order unity, the energy density 
of dark energy takes over and starts to dominate, while the other energy densities decay away.

In Figures 1 and 2 we plot the evolution of the density parameter $\Omega_i = \rho_i/\rho_{\rm cr}$ ($\rho_{\rm cr}$ is 
the critical density) of the matter species for different choices of 
$\beta$ and $\lambda$. Clearly, the interaction between the scalar field and the neutrinos 
significantly modifies the evolution of the neutrino density. The differences in the evolution of $\Omega_{\rm CDM}$ and 
$\Omega_\nu$ for different choices of $\beta$ and $\lambda$ are obvious. 
The neutrinos do not scale like $a^{-3}$ for  some time, since the pressure does not vanish completely 
at the beginning of the matter dominated epoch and also because of the coupling to the scalar field. 
This has important consequences for the early time integrated Sachs--Wolfe (ISW) effect, as we will discuss 
shortly. 
\begin{figure}
\centerline{\scalebox{0.35}{\includegraphics{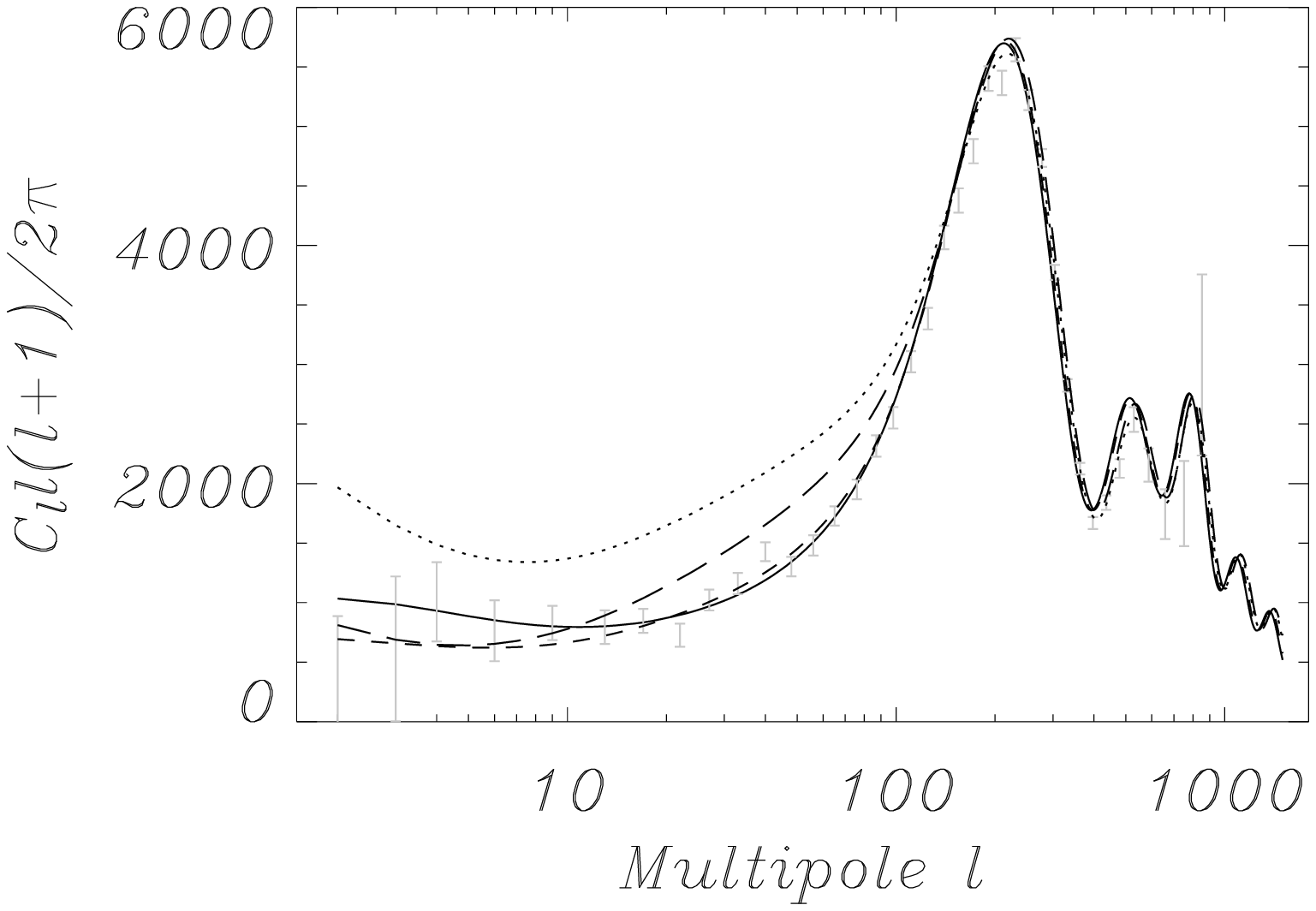}}}
\centerline{\scalebox{0.35}{\includegraphics{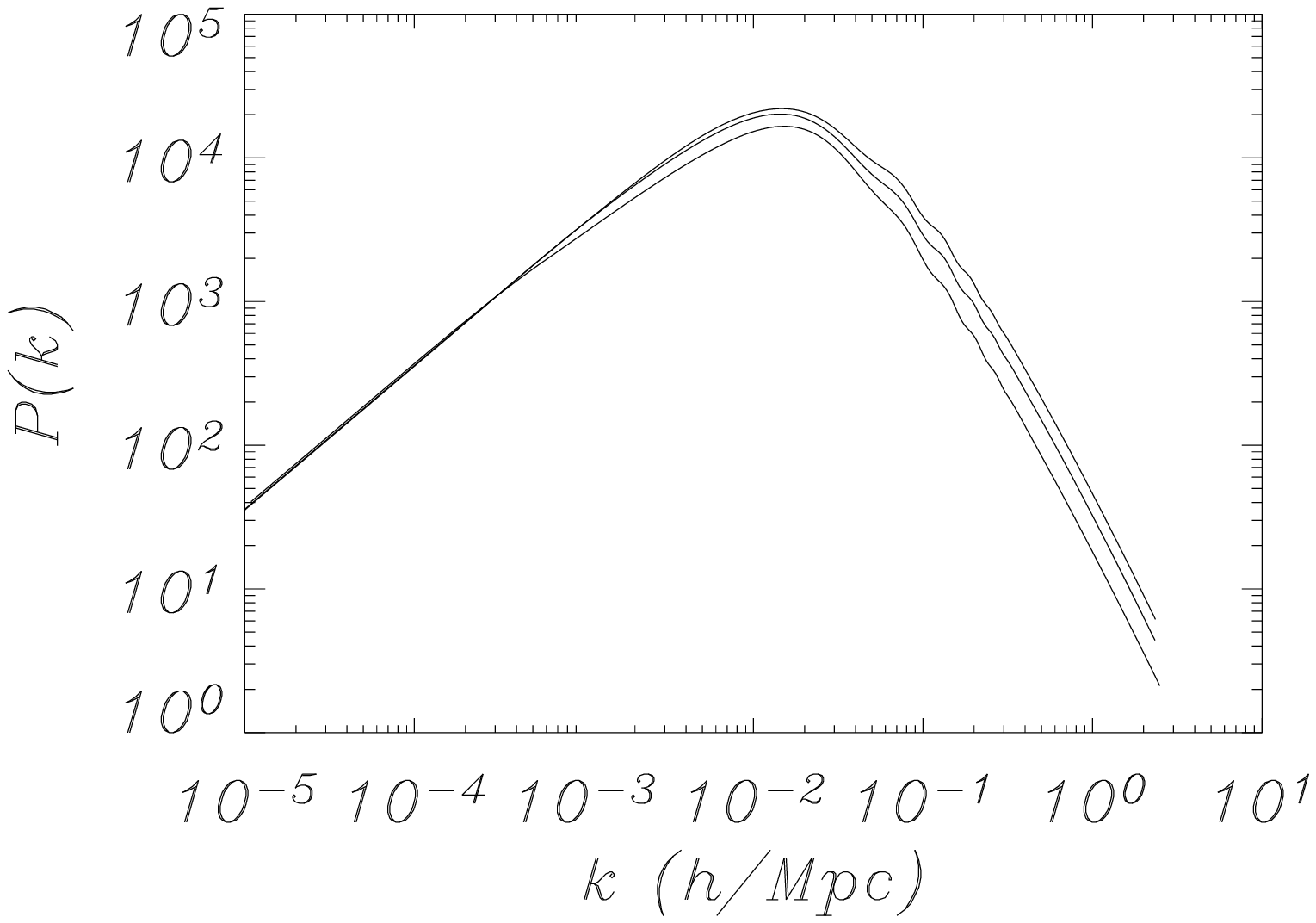}}}
\caption{Upper panel: the CMB anisotropy spectrum (unnormalized). Solid line: $\beta=0$, $\lambda = 1$; short--dashed 
line: $\beta = 1$, $\lambda=1$; dotted line: $\beta=-0.79$, $\lambda = 1$; long--dashed line: $\beta=1$, 
$\lambda=0.5$. The lower panel shows the matter power spectrum. From the top curve to the bottom curve: 
($\beta=0$, $\lambda=1$), ($\beta=1$, $\lambda=0.5$), ($\beta=-0.79$, $\lambda=1$). The matter power spectrum 
for ($\beta=1$, $\lambda=1$) is indistinguishable from the ($\beta=0$, $\lambda=1$) curve.}
\end{figure}
\begin{figure}
\centerline{\scalebox{0.35}{\includegraphics{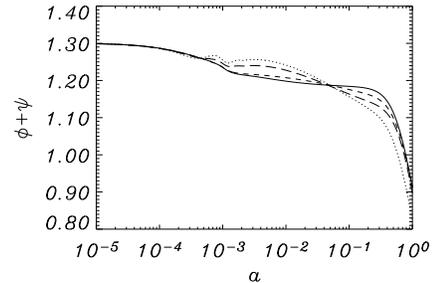}}}
\caption{Evolution of the sum of the metric perturbations $\Phi+\Psi$.
Solid line: $\beta=0$, $\lambda = 1$; short--dashed 
line: $\beta = 1$, $\lambda=1$; dotted line: $\beta=-0.79$, $\lambda = 1$; long--dashed line: $\beta=1$, 
$\lambda=0.5$.
The scale is $k=10^{-3}$Mpc$^{-1}$.}
\end{figure}

For the model studied here, the neutrinos are heavier in the past and become lighter during the cosmological
evolution. This results in a larger neutrino density in the past, for the times when the neutrinos are non--relativistic. 
Since the mass of the neutrinos is considerably larger in some models, this affects the freestreaming length 
and hence the matter power spectrum (see Figure 3). In Figure 2 (lower panel) we have also plotted the evolution of the neutrino 
mass for different choices for $\beta$ and $\lambda$. In models with positive $\beta$ the effective potential possesses 
a minimum, which explains the late time increase of the masses for the cases with $\beta>0$: the field rolls from large 
field values towards the minimum, overshoots, comes to a halt and is currently rolling back towards the minimum. 
In this case, starting from about redshift unity, the neutrino energy density will decay more slowly than in the 
cases where the minimum does not exist. This particular behaviour plays a role in the late time ISW effect (see below). 

In order to study the evolution of cosmological perturbations, we have extended the Boltzmann treatment of \cite{Ma} 
to include the coupling between neutrinos and dark energy \cite{ourlater} (see e.g. \cite{dodelsonbook} for detailed 
discussions on the physics of CMB anisotropies). To calculate the power spectra, we modified the 
CAMB code \cite{camb} accordingly. In Figure 3 we plot the anisotropy spectrum for different choices of $\beta$ and $\lambda$. 
We observe a number of differences with respect to the uncoupled case. Firstly, we see an increase in power on scales larger than a 
degree (multipole number $l<100$). In some cases 
an interesting reduction of power can be observed on larger scales (multipole number $l<10$). Furthermore, for some choices of parameters, 
the positions as well as the relative heights of the peaks are also affected. We will now discuss these effects in more detail. 

As discussed above, the background evolution is modified in the presence of mass--varying neutrinos. In particular the density 
of neutrinos is larger at early times in models with $\beta\neq 0$. Around the period of matter-radiation equality, the coupling of the neutrinos to the scalar field causes the neutrino density to decay faster than the energy density of CDM (even if the pressure of the neutrinos is negligible). This can be seen from eqn. (\ref{neutrinoequation}) and using the fact that during this period $\beta\dot \phi<0$ for
the models under consideration. As a result, the regime between the radiation 
and matter dominated era is prolonged, which can be seen in Figures 1 and 2. This implies that the evolution of the metric perturbations $\delta g_{00} = -2 a^2 \Psi$ and 
$\delta g_{ij}=-2a^2\Phi \delta_{ij}$ in this period is significantly 
modified, as shown in Figure 4. The integrated Sachs--Wolfe effect (ISW) is an integral of $\dot \Psi+\dot\Phi$ over conformal time 
and wavenumber $k$ and 
therefore depends on the parameters $\beta$ and $\lambda$. The changes in the evolution of $\Psi$ and $\Phi$ in the redshift range
$z\approx 50-1000$ imply an excess of power in the CMB spectra, which can be seen in Figure 3 in the 
multipole range $10 < l < 100$ for models with $\beta \neq 0$. 

The anisotropies on very large scales ($l \leq 20$) are dominated by the late time ISW, i.e. by the evolution of $\Psi+\Phi$ 
in the redshift range between $z=0$ and  $z\approx 1$, which is governed by the evolution 
of the background and the perturbations. In particular, $\rho_\phi$ and $\rho_\nu$ 
as well as the equation of state of dark energy affect the late time behavior of cosmological perturbations. 
As mentioned above, the evolution of the scalar field is influenced by the presence of a coupling to the neutrinos and hence the 
equation of state of dark energy depends upon $\beta$. Likewise, the clustering properties of dark energy depends on the coupling 
to neutrinos (see \cite{weller} for a discussion on the clustering of dark energy and its impact on the CMB). 
The neutrinos will generally tend to fall into the potential wells of dark matter, although at a rate slightly dependent on the 
coupling to the scalar field. The scalar field itself will cluster together with the 
neutrinos and thereby affecting the gravitational potential. For some choices of $\beta$ and $\lambda$ we find a suppression of power
relative to the case with $\beta=0$. In particular, the anisotropy spectrum for the model with $\beta=1$ and $\lambda=1$ only 
differs from the uncoupled case on large angular scales, leaving the acoustic peaks almost unmodified, and should therefore result 
in an improved fit with the latest WMAP data \cite{wmap}. However, the reduction 
of power on large angular scales is not generic and other choices for $\beta$ and $\lambda$ lead to an enhancement of power 
in the region $l=2 - 100$, as can be seen in Figure 3.

Finally, the shifts and slight rescaling in the peaks is caused by the different densities stored in massive neutrinos, baryons and CDM 
at the time of decoupling when changing the parameters $\beta$ and $\lambda$. 
The physics is very similar to the cases studied in \cite{hotmatter}. The predicted matter power spectra 
look very similar to standard models with CDM + hot dark matter. The damping observed in the spectra can be simulated by an averaged 
neutrino mass in the models considered here. However, there are new signatures in the CMB power spectra which cannot be obtained with 
an averaged neutrino mass and are due to the coupling between dark energy and neutrinos. 

We would like to point out that the decay of neutrinos into $\phi-$quanta does not play a role for the parameters 
chosen here. Potentially, this can have an important effect in cosmology (see \cite{neutrinoless}). The Lagrangian for the 
neutrinos is 
\begin{equation}
{\cal L}_\nu = m_\nu(\phi) \nu\bar\nu \approx M_0\bar{\nu}\nu + \frac{\beta M_0}{M_{Pl}} (\phi -\phi_{\rm min}) \bar{\nu}\nu + ...,
\end{equation} 
where we used $m_\nu(\phi) = M_0 \exp(\beta\phi)$, expanded around the minimum $\phi_{\rm min}$ and 
have neglected higher order terms. This Lagrangian has the same form as the one used in \cite{neutrinoless} 
if we identify the coupling to be $g = \beta M_0/M_{Pl}$. For $\beta={\cal O}(1)$ and $M_0 \approx {\cal O}({\rm eV})$
the coupling $g \ll 1$ and indeed much smaller than the value ($g=10^{-5}$)
used in \cite{neutrinoless}.

In conclusion, cosmologies with neutrino--dark energy coupling have a rich phenomenology. It is clear from the results 
presented in this letter that models with mass--varying neutrinos cannot be mimicked with an averaged constant neutrino 
mass. We have found that some models with a coupling of the order of the inverse of the Planck mass present a 
reduction of power in the temperature CMB anisotropies spectrum at low multipoles but a standard cosmology peak structure 
in line with current CMB data. Our work implies that CMB anisotropies as well as large scale structures will be able to 
constrain parameters of the theory tightly. In future we will investigate other potentials and couplings, elaborate on the 
degeneracies between the parameters and will use current data to constrain such models \cite{ourlater}. 

We are grateful to O. Elgaroy, D. Hooper, A. Melchiorri, J. Silk and C. Skordis for useful discussions and to 
A. Lewis for allowing us to use the CAMB quintessence module. The authors 
were supported by PPARC (AWB) and the Research Council of Norway through project number 159637/V30 (DFM). 
DTV acknowledges a Scatcherd Scholarship. 
\end{document}